%% file: conference_041818.tex
\documentclass[conference]{IEEEtran}
\IEEEoverridecommandlockouts

\usepackage{cite}
\usepackage{amsmath,amssymb,amsfonts}
\usepackage{algorithmic}
\usepackage{graphicx}
\usepackage{textcomp}
\usepackage{xcolor,mathcomSTEv4}
\def\BibTeX{{\rm B\kern-.05em{\sc i\kern-.025em b}\kern-.08em
    T\kern-.1667em\lower.7ex\hbox{E}\kern-.125emX}}

\usepackage{enumerate}
\usepackage[small,bf]{caption}
\usepackage{dsfont,pgfplots}
\usepackage{bbm, mathcomSTEv4}
\usepackage{tikz}
\usepackage{mathabx}
\usetikzlibrary{positioning}
\usetikzlibrary{arrows.meta}
\usetikzlibrary{decorations.markings}
\usetikzlibrary{decorations.pathreplacing,decorations.markings}
\usetikzlibrary{arrows,calc,decorations.markings,math,arrows.meta}

\usepackage{graphicx,psfrag,amsmath,amsfonts,verbatim,amsthm, amssymb}
\usepackage{enumerate}

\usetikzlibrary{positioning}
\usetikzlibrary{arrows.meta}
\usetikzlibrary{decorations.markings}
\usetikzlibrary{decorations.pathreplacing,decorations.markings}
\usetikzlibrary{arrows,calc,decorations.markings,math,arrows.meta}
\usetikzlibrary{shapes.geometric}

\tikzstyle{int}=[draw, fill=blue!10, minimum height = 1cm, minimum width=1.5cm,thick ]

\pgfplotsset{compat=1.17}

\begin{document}

\title{An Exploration of the\\Heterogeneous Unsourced MAC}

\author{\IEEEauthorblockN{
\dag A. Hao,
\dag S. Rini,
\S V. K. Amalladinne,
\S A. K. Pradhan, 
\S J.-F. Chamberland \\
\dag Electrical and Computer Engineering, National Chiao Tung University \\
\S Electrical and Computer Engineering, Texas A\&M University
}
\thanks{This material is based upon work supported, in part, by the National Science Foundation (NSF) under Grant CCF-1619085 and by Qualcomm Technologies, Inc., through their University Relations Program.}
}

\maketitle

\begin{abstract}
The unsourced MAC model was originally introduced to study the communication scenario in which a number of devices with low-complexity and low-energy wish to upload their respective messages to a base station.
In the original problem formulation, all devices communicate using the same information rate.
This may be very inefficient in certain wireless situations with varied channel conditions, power budgets, and payload requirements at the devices. 
%
This paper extends the original problem setting so as to allow for such variability.
More specifically, we consider the scenario in which devices are clustered into two classes, possibly with different SNR levels or distinct payload requirements.
In the cluster with higher power, devices transmit
using a two-layer superposition modulation.
%
%
In the cluster with lower energy, users transmit with the same base constellation as in the high power cluster.
Within each layer, devices employ the same codebook.
At the receiver, signal groupings are recovered using Approximate Message Passing (AMP), and proceeding from the high to the low power levels using successive interference cancellation (SIC).
This layered architecture is implemented using Coded Compressed Sensing (CCS) within every grouping.
An outer tree code is employed to stitch fragments together across times and layers, as needed.
This pragmatic approach to heterogeneous CCS is validated numerically and design guidelines are identified.
\end{abstract}


\section{Introduction}

The IoT paradigm of myriad unattended devices connected wirelessly to the Internet may pose a significant disruption to existing communication networks.
The predicted number of such devices, orders of magnitude greater than human subscribers, and the usage profile of these devise, sporadic and fleeting, invalidate the type of connection-based architectures that form a foundation for existing deployments.
Thus, new means of Internet access must be explored to reflect this change, with provisions for random access.
Along these lines, one model attuned to this reality that has gained attention in recent years is unsourced random-access (URA).
The URA formulation, originally proposed by Polyanskiy~\cite{polyanskiy2017perspective}, centers on concurrent up-link data transfers.
There is a strong connection between URA and Compressed Sensing (CS), with the former problem being an instance of a noisy support recovery task~\cite{choi2017compressed,reeves2012sampling,gilbert2017all}.
More precisely, in URA, the receiver seeks to identify the set of messages being transmitted by active devices, without regard for the identities of their sources.
The identity of a source can be embedded in the message payload, if needed.
The value of this approach lies in the fact that the access point does not need to determine the set of active devices at the onset of a frame, a step that can rapidly become overwhelming for connection-less settings with a very large population of candidate transmitters.
%
URA raises both theoretical and practical challenges.
Achievable bounds rooted in finite-block length analysis for such systems can be found in~\cite{polyanskiy2017perspective}.
These bounds are obtained devoid of complexity constraints, as they rely on joint maximum likelihood decoding.

Several pragmatic, low-complexity approaches for this problem have been proposed~\cite{ordentlich2017low,vem2019user,amalladinne2019coded,calderbank2018chirrup,fengler2019sparcs,pradhan2019sparseidma,marshakov2019polar,AKPolar}.
Conceptually, each of these contributions offer a means to circumvent the difficulty associated with the dimensionality of the problem.
Indeed, when viewed as a support recovery task, unsourced random access features a $K$-sparse state vector of length $2^{100}$ or longer.
This reality prevents the straightforward application of standard CS solvers.
To address this issue, many algorithms leverage lessons from random access and coding theory to design structured sensing matrices suitable for the efficient recovery of the sent messages.
%
A line of research that has attracted attention in this context is the framework of coded compressed sensing (CCS) originally proposed by Amalladinne et al.~\cite{amalladinne2018couple,amalladinne2019coded}.
This scheme is a divide-and-conquer approach where a large CS problem is partitioned into smaller components, each of which can be solved using standard CS algorithms.
The output of this step produces lists of message fragments, one list for every CS instance.
The transmitted messages are recovered by stitching fragments together using an outer code.
Overall the approach can be abstracted as a concatenated coding scheme where an inner code is task with fragment recovery and the outer code is responsible for message disambiguation.

CCS has been ported, enhanced, and extended by multiple authors.
It appears as a component of the ultra-low complexity CHIRRUP algorithm~\cite{calderbank2018chirrup}, and it can be incorporated into activity detection in multi-antenna systems~\cite{fengler2019massive}.
CCS can be employed to build neighbor discovery schemes and to handle signal asynchrony~\cite{zhang2013neighbor,thompson2018compressed,amalladinne2019asynchronous}.
An enhanced version of the algorithm takes advantage of the fact that output from the early stages of CCS can be integrated into later stages as side information to improve execution~\cite{amalladinne2020enhanced}.
This variant has inspired significant extensions related to sparse regression codes and Approximate Message Passing (AMP)~\cite{fengler2019sparcs,amalladinne2020unsourced}.


Along similar lines of research, the main contributions of our article can be summarized as follows.
\begin{itemize}
\item In Sec.~\ref{sec:Heterogeneous System Model}, we introduce a novel system model for unsourced random access.
This new model captures the fact that, in practice, wireless IoT devices may have distinct payload requirements.
Heterogeneity is addressed by introducing the notion of clustering, whereby users within a cluster have the same power budget and they transmit at the same rate.
We refer to this model as HetURA.
\item A pragmatic communication scheme for this setting is developed in Sec.~\ref{sec:A Coded Compress Sensing  Scheme}.
The propose algorithm borrows ideas from CCS~\cite{amalladinne2018couple,amalladinne2019coded}, but also introduces a phased decoding approach akin to successive interference cancellation across layers.
Portions of clusters with greater energy budgets are decoded first.
The structure of the problem is facilitated by a superposition constellation with two-levels.
\end{itemize}
The value of the proposed framework is examined in Sec.~\ref{sec:Numerical Evaluations}, where performance results showcase the validity of the approach.
Finally, Sec.~\ref{sec:Conclusion} concludes the paper. 





\section{Heterogeneous URA}
\label{sec:Heterogeneous System Model}

\begin{figure*}
    \centering
    \input{HetURA}
    \caption{This notional diagram offers a synopsis of the proposed communication scheme, as described in Sec.~\ref{sec:A Coded Compress Sensing  Scheme}.}
    \label{fig:scheme}
\end{figure*}
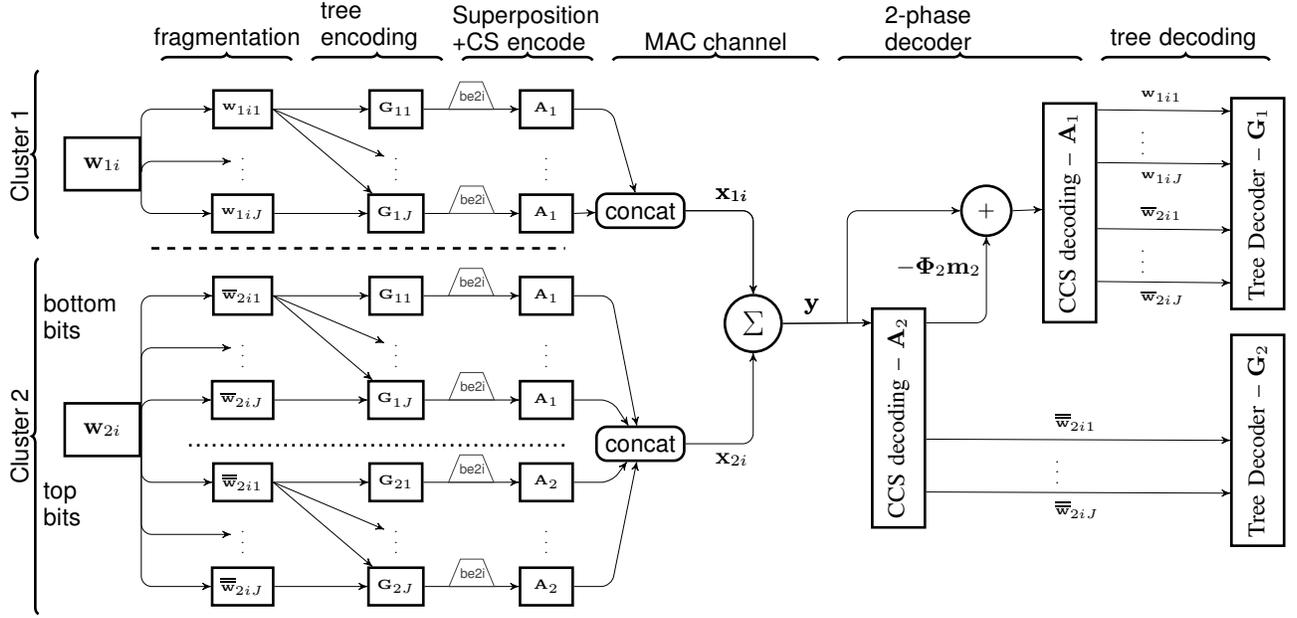
Our goal is to introduce and study a heterogeneous version of URA with groupings, where distinct groups have different power levels and data requirements.
We refer to this model as the heterogeneous URA (HetURA).
%
%
%
The HetURA is formally defined as the up-link scenario where the user population is divided in $K$ clusters, with cluster~$k$ containing the set of devices $\mathcal{S}_k$.
Of these $\mathcal{S}_k$ devices, only a subset $\mathcal{A}_k \subset \mathcal{S}_k$ of size $|\mathcal{A}_k|=M_k$ is active, with users therein wishing to communicate to the base station.
The output at the receiver is then equal to
\ea{
\yv = \sum_{k \in [K]}  \sum_{i \in \mathcal{A}_k} \xv_{ki} + \zv,
\label{eq:yv N}
}
where $\xv_{ki} \in \Rbb^N$ is the channel input of the $i^{\mathrm{th}}$ user in the $k^{\mathrm{th}}$ cluster, and $N$ is the block-length. Note $[K] \ \triangleq  \{1,\ldots,K\}$ in \eqref{eq:yv N}.
Each input sequence is subject to an expected power constraint $\|\xv_{ki}\|_2^2 \leq N P_k$ with cluster ordering $P_k \leq P_{k+1}$.
The components of additive noise $\zv$ are independent, each with a standard normal distribution.

A suitable transmission scheme for the HetUMAC is defined as follows.
Active user~$i$ in cluster~$k$ wishes to transmit message $\wv_{ki} \in \lsb \lfloor 2^{N R_k} \rfloor \rsb$, where $R_k$ denotes the rate of cluster~$k$.
All the users within a cluster employ the same code and, hence, they share a same rate.
This gives
\ea{
 \xv_{ki}=f_{{\rm enc}-k} \lb \wv_{ki}\rb, \quad \forall \ i \in \mathcal{A}_k.
\label{eq:encoding function}
}
Having observed  $\yv$, the receiver is tasked with decoding the list of messages transmitted by each cluster; that is,
%
\eas{
\Wcal_k = f_{{\rm dec}-k}(\yv),  \quad \forall \ k \in [K], 
}
with $|\Wcal_k|=M_k$.
Every entry on this list should take value in the set $\lsb \lfloor 2^{N R_k} \rfloor \rsb$.
System performance is evaluated according to the per-user probability of error, defined as~\cite{polyanskiy2017perspective}
\ea{
P_{\rm UE} = \max_{k \in [K]} \ \f 1 {M_k} \sum_{i \in \mathcal{A}_k} \Pr \lsb  w_{ki} \not \in  \Wcal_k | \yv \rsb .
}
In words, this captures the (maximum) probability that a message sent by one of the devices is not recovered at the receiver.
%
%
Note that, since all the users in a cluster use the same encoding function, as in \eqref{eq:encoding function}, the receiver does not discover which user transmitted which message.
%
%
%
%

\section{A Coded Compressed Sensing Scheme}
\label{sec:A Coded Compress Sensing  Scheme}
In this section, we describe an extension of the work found in~\cite{amalladinne2019coded} adapted to the HetUMAC scenario discussed in Sec.~\ref{sec:Heterogeneous System Model}. 
To put our contribution in context, we begin with a brief review of key CS notions.
%
%
The original URA formulation can be viewed as sparse support recovery from observation 
\ea{
\yv = \Phim \mv + \zv,
\label{eq:CCS basic}
}
where $\Phim \in \Rbb^{N \times 2^{\lfloor N R \rfloor}}$ is a dictionary of possible signals, $\mv$ is a binary vector that contains the indices of the transmitted codewords so that $\mv \in \{0,1\}^{2^{\lfloor N R \rfloor}}$, and $\zv$ is additive noise as in \eqref{eq:yv N}.
We stress that $\mv$ is a sparse vector with $\|\mv \|_0$ being equal to the number of active URA devices.

As mentioned above, this article explores the extended scenario where the device population is partitioned into groups, and users from distinct clusters employ different codebooks.
For ease of exposition, we restrict our treatment to the case where $K=2$.
%
When two groupings are present, the CS interpretation of URA becomes
\ea{
\yv =  \Phim_1 \mv_1  + \Phim_2  \mv_2 + \zv.
\label{eq:CCS 2}
}
In a manner analogous to the basic URA formulation, $\mv_1$ denotes the collection of indices from the first cluster; and $\mv_2$, the indices from the second cluster. 
As in Sec.~\ref{sec:Heterogeneous System Model}, we assume that the clusters are ordered in increasing transmit power, so that we refer to the first/second cluster and low/high-energy cluster.
For simplicity, we do not discuss this alternative in the paper.
%
%
Recall that CCS was introduced as a means to tackle the dimensionality issue posed by the width of $\Phim$.
Quite obviously, when expanding the sensing matrix $\big[ \Phim_1 \, \Phim_2 \big]$ to accommodate multiple groupings, a number of complexity issue arises.
In particular, similarly to CSS, complexity allows for the decoding through non-negative least squares (NNLS) or LASSO only for limited problem dimensions in \eqref{eq:CCS 2}.
To support longer transmission block-lengths, the transmitted bits are divide into fragments and sent in separate slots. 
Since the identity of the transmitter is not conveyed in the choice of encoding function, the individual fragments of the original messages must be pieced together through a low-complexity tree-based algorithm as in \cite{amalladinne2019coded}.

To further reduce decoding complexity, the users in the high-energy cluster transmit their message bits using a superposition constellation with two-layers: a top and a bottom layer.
The symbols in the bottom layer are transmitted using the same code-book as the low-energy cluster, whereas the remaining bits are ``on top'' of the bottom bits using superposition constellation. 
%
%
%
%
%
This coding choice allows the receiver, for each transmission slot, to first decode the top layer of the superposition constellation in the high-energy cluster; and then, after using Successive Interference Cancellation (SIC), decode the low-energy users together with the bottom layer in the superposition constellation of the high-energy cluster. 
Upon decoding all the message fragments in all the slots and from all the users, the receiver can then employ the tree decoder to reconstruct the set of transmitted messages. 
We further detail the proposed transmission scheme below.

\subsection{Fragmenting}
In the low-energy cluster, every message $\wv_{1i}$ is converted into a binary vector and partitioned into $J$ sub-blocks, where the $j^{\mathrm{th}}$ sub-block consists of $B_{1j}$ bits, so that $\sum_{j \in [J]} B_{1j}=B_1=\lfloor N R_1 \rfloor$. 
This results in a collection of information fragments $\{\wv_{1ij}\}_{j\in [J]}$ for message $\wv_{1i}$.
%
On the other hand, users in the high-energy cluster first split their bits into two groups: one for the top layer and one for the bottom layer.
Denote these two sets of bits by $\woov_{2i}$, $\wov_{2i}$ for the top and bottom portions, respectively.
Likewise, let $\Boo_{2}$, $\Bo_{2}$ be the total numbers of bits assigned to these two layers; and $\Roo_{2}$, $\Ro_{2}$ be the corresponding rates.
Subsequently, the bottom bits  are fragmented exactly as in the low-energy cluster to form the set  $\{\wov_{2ij}\}$.
The top bits $\woov_{2i}$ are also partitioned into $J$ fragments, but this time $\Boo_{2j}$ is the size of the $j^{\mathrm{th}}$ fragment (not necessarily the same partitioning as in the bottom layer). Again, we must have $\sum_{j \in [J]} \Boo_{2j}=\Boo_2$ and, additionally, $\Bo_2+\Boo_2=\lfloor N R_2\rfloor$.

\subsection{Tree Encoding}
\label{sec:Tree Encoding}
The role of the tree decoder is to enable the stitching of message fragments at the decoder.
In CCS, this is accomplished by appending parity bits to the $j^{\mathrm{th}}$ fragment based on the preceding information bits.
Every device in the low-energy cluster takes the message fragment $\wv_{1i(j+1)}$ and encodes it into a vector $\vv_{1i(j+1)}$ using a systematic random linear code, together with the message fragment, $\wv_{1i(j+1)}$
\ea{
\vv_{1i(j+1)}=\lsb \wv_{1i(j+1)} \ ; \   \Gv_{1j} \otimes \lsb \wv_{1i1} ; \ldots ; \wv_{1ij} \rsb \rsb,
\label{eq:tree out 1}
}
where $\Gv_{1j}$  
produces $T-B_{1(j+1)}$ random parity bits from all the previous segments, and $\otimes$ indicates modulo-2 matrix multiplication.
Thus, $\vv_{1i(j+1)}$ is viewed as a binary vector.
In this scheme, $B_{11}=T$; that is the first fragment does not contain any parity bits.
%
%
%
%
In \eqref{eq:tree out 1}, $T \geq B_{1j}$, so that effectively we have $T-B_{1j}$ random linear parity constraints embedded in this block to help stitch together fragments of information bits belonging to $\wv_{1ij}$ when decoding codeword $\vv_{1i}$. 
%
%

A user in the high-energy cluster performs a similar encoding process.
Redundancy for bottom bits is added paralleling the encoding in the low energy cluster, yielding
\ea{
\vov_{2i(j+1)}=\lsb \wov_{2i(j+1)}  \ ; \  \Gv_{1j} \otimes \lsb \wov_{2i1} ; \ldots ; \wov_{2ij} \rsb \rsb,
\label{eq:tree out 2}
}
where $\Gv_{1j}$ is the same binary matrix that appears in \eqref{eq:tree out 1}.
The top bits are encoded according to the information bits in each cluster, i.e.,
\ea{
\voov_{2i(j+1)}= \lsb \woov_{2i(j+1)} \  ; \ \Gv_{2j}  \otimes \lsb \woov_{2i1} ; \ldots ; \woov_{2ij} \rsb  \rsb,
}
where $\Gv_{2j}$ is, again, a random parity generating matrix.

\subsection{Superposition Coding \& CS Encoding}
After tree encoding is complete, each encoded block has size $T$.
These blocks are then encoded using two set of inner CS codes: (i) one for the segments of the low-energy cluster and the bottom segments of the high-energy cluster; and (ii) one for the top segments of the high-energy users.
To apply the inner encoding, we convert the binary string $\vv_{1ij}$ into the one-norm binary vector  $\mv_{1i} \in \{0,1\}^{2^T}$ in which a single one is placed at the location corresponding to the integer value of $\vv_{1ij}$.
This is the emblematic index representation used in CCS.
Blocks $\vov_{2ij}$ and $\voov_{2ij}$ are converted to their index representations in a similar manner.

The two CS code differ as follows: the former has entries from $\{+\sqrt{P_1},-\sqrt{P_1}\}$, while the latter features entries from $\{+\sqrt{P_2-P_1},-\sqrt{P_2-P_1}\}$.
This difference in support results in a superposition constellation; the top bits are effectively transmitted at a higher power than the bottom bits, based on our assumption $P_2 > 2P_1$.
Accordingly the CSS signal corresponding to section $\vv_{1ij}$ is
\ea{
\xv_{1ij} = \Av_1 \mv_{1ij},
\label{eq:fragment channel input 1}
}
where $\Av_1 \in \{+\sqrt{P_1},-\sqrt{P_1}\}^{ Q  \times 2^T }$  is a matrix formed by picking $Q=\lfloor N R / J \rfloor$ rows uniformly at random (excluding the row of all ones) from a Hadamard matrix of dimension $2^T \times 2^T$ and re-scaling them to meet the power constraint.
%
Similarly, after tree encoding, the bottom bits are CS encoded,
\ea{
\xov_{2ij} = \Av_1 \mov_{2ij}
\label{eq:fragment channel input 2 bottom}
}
using the same sensing matrix.
The top bits are processed using a different signal dictionary,
\ea{
\xoov_{2ij} = \Av_2 \moov_{2ij} .
\label{eq:fragment channel input 2 top}
}
The rows of matrix $\Av_2 \in \{+\sqrt{P_2-P_1},-\sqrt{P_2-P_1}\}^{ Q  \times 2^T }$ are also scaled versions of randomly selected rows from a Hadamard matrix (excluding the row of all ones).

Finally, the channel inputs are obtained by concatenating the partial signals.
For the low-energy users, we get
\ea{
\xv_{1i} & = \lsb \xv_{1i1}  \ \xv_{1i2} \ \ldots \    \xv_{1iJ} \rsb,
\label{eq:concatenate x}
}
and, for the high energy users, we have
\eas{
\xov_{2i} & = \lsb \xov_{2i1}  \ \xov_{2i2} \ \ldots \ \xov_{2iJ} \rsb \label{eq:concatenate x2o} \\
\xoov_{2i} & = \lsb \xoov_{2i1}  \ \xoov_{2i2} \ \ldots \  \xoov_{2iJ} \rsb
\label{eq:concatenate x2oo}\\
\xv_{2i} \   & = \xov_{2i}+  \xoov_{2i},
}
Over the ensemble of random coding parameters, we obtain expected transmit power
\begin{equation*}
\begin{split}
\left\| \xv_{2i} \right\|^2 &= \left\| \xov_{2i} \right\|^2 + \left\| \xoov_{2i} \right\|^2 \\
&= N P_1 + N (P_2 - P_1) = N P_2 ,
\end{split}
\end{equation*}
as mandated by the constraint associated with \eqref{eq:yv N}.
The parameters of the scheme are summarized in Table.~\ref{tab: recap}.

\subsection{Channel Transmission/Reception}
As the transmitted channel inputs are composed of coded segments of the same length, we observe that the CCS construction naturally maps to the CS interpretation in \eqref{eq:CCS 2}.
To see this, let $\mv_{1j}$ be the sum of coded fragments $\mv_{1ij}$ for $i \in \mathcal{A}_1$, then let $\mv_{1}$ be the concatenation of the fragments  $\mv_{1j}$ as in \eqref{eq:concatenate x}.
Define $\mov_{2j}$, $\mov_{2}$ and $\moov_{2j}$, $\moov_{2}$ in a similar manner.
Let $\Phim_1$ be the tensor product between $\Iv_{J}$ and $\Am_1$; likewise, let $\Phim_2$ be that between $\Iv_{J}$ and $\Am_2$.
Then, we can express the received signal as
\ea{
\yv = \Phim_1 \lb \mv_1 + \mov_{2} \rb + \Phim_2 \moov_{2}+\zv.
\label{eq:out 2}
}
The signal aggregate obtained by adding $\mv_1$ and $\mov_{2}$ has a sparsity of $J(M_1 + M_2)$, 
whereas $\moov_{2}$ is $J M_2$ sparse.

\subsection{Two-Phase Decoding}
\label{sec:Two-Phase Decoding}
Upon getting observation $\yv$, the receiver separates it into sections $\{\yv_j\}_{j\in [J]}$, each block corresponding to the summation of the sections $\Av_1 \mv_{1j}$, $\Av_1 \mov_{2j}$, and $\Av_2 \moov_{2j}$, as in \eqref{eq:out 2}. 
The receiver begins by decoding $\moov_{2j}$ using the CSS algorithm with coding matrix $\Av_2$.
During this phase of the decoding process, it treats the remaining terms in $\yv_j$ as additional noise.
Parameters are selected to make sure that this portion of the decoding process is successful with high probability.
Furthermore, for ease of exposition, we assume that $\moov_{2j}$ is correctly recovered, although in reality, an error at this stage will produce some interference at the subsequent stage.

Once $\moov_{2j}$ is recovered, the decoder computes a residual, or effective observation, for every section,
\begin{equation*}
\tilde{\yv}_j = \yv_j - \Av_2 \moov_{2j} .
\end{equation*}
In the spirit of successive interference cancellation, these sections, $\left\{ \tilde{\yv}_j \right\}_{j\in [J]}$ are then passed to the CSS algorithm with coding matrix $\Av_1$ and sparsity level $M_1 + M_2$.

The recovery algorithm we adopt for individual sections is an AMP-based CS solver~\cite{maleki2010optimally}.
Such composite algorithms iterate through two equations:
\begin{align}
\begin{split}
\zv^{(t)} &= \yv - \Am \mv^{(t)} \\
&\quad + \frac{\zv^{(t-1)}}{Q} \operatorname{div} \boldsymbol{\eta}_{t-1} \left( \Am^{\mathrm{T}} \zv^{(t-1)} + \mv^{(t-1)} \right)
\end{split} \label{equation:AMP00} \\
\mv^{(t+1)} &= \boldsymbol{\eta}_t \left( \Am^{\mathrm{T}} \zv^{(t)} + \mv^{(t)} \right),
 \label{equation:AMP01}
\end{align}
with initial conditions $\mv^{(0)} = \zerov$ and $\zv^{(0)} = \yv$.
The function $\boldsymbol{\eta}_t (\cdot)$ in \eqref{equation:AMP01} is the denoiser, which can take the form of a posterior mean estimate~\cite{Giuseppe} or a standard soft thresholding operator~\cite{daubechies2004iterative,beck2009fast}.
Equation \eqref{equation:AMP00} can be interpreted as the computation of a residual enhanced enhanced with an Onsager correction~\cite{bayati2011dynamics,donoho2013information}.

After converting the terms  $\mv$ into $\vv$, the tree decoder uses the random parity bits to stitch together the information bits from each of the users, thus reconstructing the transmitted terms $\wv_1$ and $\wv_2$.\footnote{In actuality, one also needs some random parity bits to stitch together the top and the bottom bits from the high-energy users. For simplicity this coding step is not presented here, as it is analogous to the step in Sec.~\ref{sec:Tree Encoding}.}

The encoding and decoding process is also conceptually represented in Fig. \ref{fig:scheme}.
Successive steps in the scheme  are represented in vertical sections, proceeding from left to right. 
Separated horizontal section represent the processing of three sets of bits: the bits from cluster~1, the bottom bits from cluster~2, and the top bits from cluster~2.
In the figure ``be2i'' indicates the conversion from binary string of length $l$ to the index in the $2^l$ binary vector. 
Also, ``concat'' indicates the concatenation of the segment as in \eqref{eq:concatenate x}.

\begin{table}[h]
\centering
\begin{tabular}{ |l|l||l| l | } 
 \hline
 Quantity  &   & Quantity &   \tabularnewline
 \hline
 block-length & $N$  &   \# fragments & $J$ \tabularnewline
\# clusters  & $K$ & len. low-energy fragments  & $B_{1j}$ \tabularnewline
\# active users & $M_k$ & len.  high-energy top fragments &   $\Bo_{2j}$ \tabularnewline
power constraint & $P_k$ & len.  high-energy top fragments &  $\Boo_{2j}$ \tabularnewline 
transmission rate & $R_k$ & len.  tree-coded fragments  & $T$  \tabularnewline
\# transmitted bits & $B_k$  &  len. CSS+tree-coded fragments & $Q$ \tabularnewline
 \hline
\end{tabular}
\caption{Summary of the quantities in Sec.~\ref{sec:Heterogeneous System Model} and Sec.~\ref{sec:A Coded Compress Sensing  Scheme}}.
\label{tab: recap}
\end{table}

\section{Numerical Evaluations}
\label{sec:Numerical Evaluations}
We now turn to the numerical simulation of the scheme introduced in Sec.~\ref{sec:A Coded Compress Sensing  Scheme}.
Generally speaking, we are interested in arguing that the scheme in Sec.~\ref{sec:A Coded Compress Sensing  Scheme} allows high-energy users to transmit at high rates while preserving the baseline performance in which all nodes, i.e. $M_1+M_2$, transmit at the low-energy level $P_1$  as in~\cite{amalladinne2019coded}.
Indeed, this is the design reasoning behind the coding choice of treating the bottom bits of the high energy cluster as the bits in the low energy cluster. 
%
Accordingly, in the simulations we fix scheme parameters in Table~\ref{tab:sim} and study the rate performance as a function of $P_1$ while $P_2/P_1=\al$ is kept fixed.
%

\begin{table}[h]
\centering
\begin{tabular}{ |l|l||l|l|} 
\hline
Parameter & Value  & Parameter & Value \\
\hline
block-length & $N = 200$  & \# of fragments & $J=11$  \\
\# low-energy users  & $M_1=10$ & section length  & $L=5$ \\  
\# high-energy users  & $M_2=5$ & \# of AMP iterations & $10$ \\ 
low-energy payload  & $B_1=100$ & target $P_{\rm UE}$ & 5\% \\
%
%
%
\hline
\end{tabular}
\caption{Summary of the simulation parameters in Sec.~\ref{sec:Numerical Evaluations}.}
\label{tab:sim}
\end{table}

\subsection{Baseline Performance}
\label{sec:Baseline-performance}
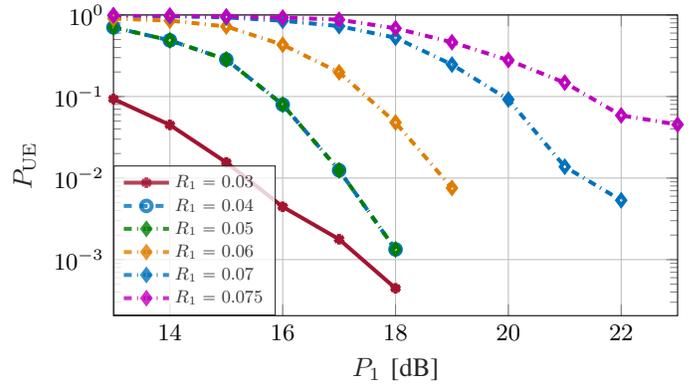
\begin{figure}
    \centering
\input{rate_saturation}
    \caption{The $P_{UE}$ for a fragment in  the baseline URA  in Sec. \ref{sec:Baseline-performance} as a function of $P_1$ for various rates $R_1 \in [0.03,0,075]$.}
    \label{fig:basic URA}
\end{figure}

In this section, we elaborate on the baseline settings for our simulation campaign. 
The  performance associated with the original CSS code in each fragment is plotted in Fig.~\ref{fig:basic URA} for the settings in  Table~\ref{tab:sim}.
%
%

Let us consider the performance of the classic (coordinated) RA scheme in which each active user transmits using time-sharing for a portion $1/(M_1+M_2)$ of the time.
In this case, the largest attainable rate (ignoring the block-length effects) is $R_1=0.067$: note the CCS code here attains reliable decoding up to   $R_l=0.07$ but shows a $P_{\rm UP}$ saturation at $R=0.08$.
%

To attain the desired block-length, the various fragment is stitched together using a tree code expressed as $\Bv_1=[0,5,5,5,5,5,5,5,5,5,9]$ where the $j^{\rm}$ element of $\Bv_1$ is $B_{1j}$ in \eqref{eq:tree out 1}.
With this choice choice of tree-coding parameters, we achieve an error rate below $1\%$ at the tree decoder.
%
%
\subsection{Comparison with TDMA}
\label{sec:Comparison with TDMA}
Next, we wish to compare the performance of the scheme in Sec. \ref{sec:A Coded Compress Sensing  Scheme} with the simpler scheme relying time-division multiple access (TDMA) as follows.
Given a block-length $N$, transmission takes place in a low-rate phase of duration $\la N$ and a high-rate phase of duration $(1-\la)N$.
In the low-rate phase, all users send at the baseline rate in Sec. \ref{sec:Baseline-performance} at power $P_1'=P_1/\la$ while, in the high-phase rate, the high-rate users transmit at rate the baseline rate for $M_2$ users at power $P_2'$ obtained as 
\ea{
P_2' (1-\la)+\la P_1= P_2.
}
Let the rate attained with this scheme in the two clusters as $(R_1',R_2')$.
We can compare the performance of this TDMA approach with the approach in Sec. \ref{sec:A Coded Compress Sensing  Scheme} by letting the rate in the low-energy user be $R_1'$ and see what rate for the high-energy user is attainable. 
We can then interpret $\Roo_2$ and $R_2'$ as the rates that the two approaches afford the high-energy users for a given degradation of the baseline performance.
%
%
%
\begin{figure}
    \centering
    \hspace{-0.5cm}
\input{Sim_one_layer_powerformance}
    \caption{Simulation results for the setting in Sec. \ref{sec:Comparison with TDMA} for $\al=P_2/P_1=6$}
    \label{fig:Sim_one_layer_powerformance}
\end{figure}
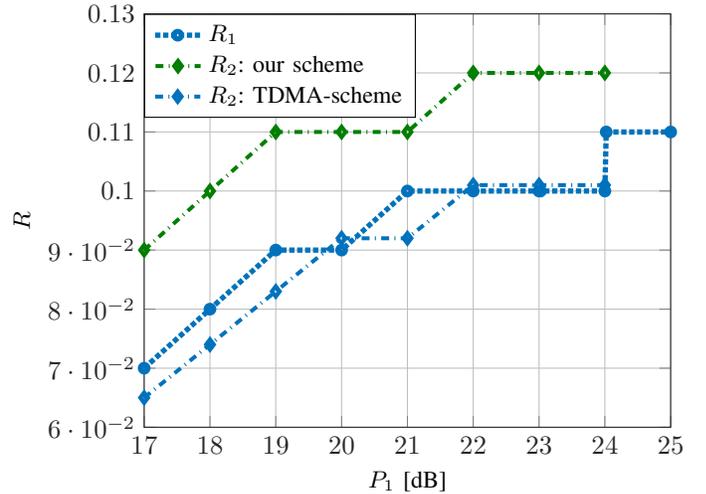

\subsection{IAN improvement}
\label{sec:SIC}
\begin{figure}
    \centering
    \hspace{-0.5cm}
    \input{SIC_CDMA}
    \caption{Decoding improvement with respect to the interference as noise (IAN) prediction described in Sec. \ref{sec:SIC}.}
    \label{fig:SIC_CDMA}
\end{figure}
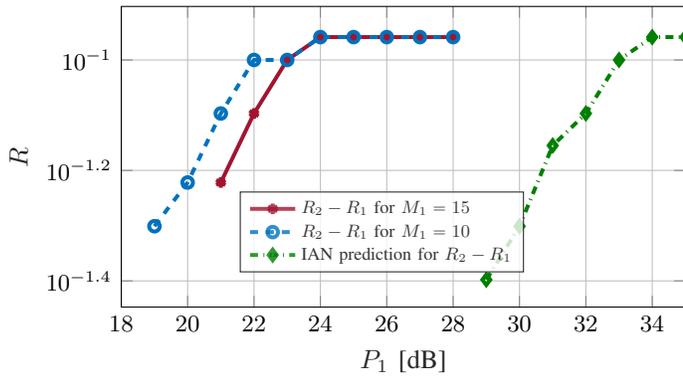

Thorough numerical experiments, we have noted that  the decoding procedure in Sec. \ref{sec:Two-Phase Decoding} actually greatly outperforms  the performance predicted by SIC assuming that the effecting noise $\Phim_1 \lb \mv_1 + \mov_{2} \rb+\zv$ is normal distributed, i.e. IAN.
One inherent feature of the CCS codes used in Sec. \ref{sec:A Coded Compress Sensing  Scheme} is that the randomly generated codewords in $\Av_1$/$\Av_2$ are uniformly distributed in the code space. 
This means that, when $R_1$ and $R_2$ are sufficiently low, the codewords in $\Av_1$ and $\Av_2$ are perpendicular with high probability.
We indeed observe that the decoding procedure in  Sec. \ref{sec:A Coded Compress Sensing  Scheme} is inherently able to exploit the codeword perpendicularity in the projection step.
We are currently unable to precisely characterize this effect; 
nonetheless, we can numerically investigate  this phenomena as in Fig. \ref{fig:SIC_CDMA}.
%
Here we plot the largest $\Roo_2=R_2-R_1$ attainable for different number of users, together with the IAN prediction. 

\section{Conclusion}
\label{sec:Conclusion}

In this paper, we consider the an extension to the unsourced random access (URA) scheme in which the active users are divided into two sets: a set of high-energy and one of low-energy users.
We propose a transmission scheme which relies on superposition constellation and successive interference cancellation in order to accommodate a large rate for the high-energy users.
This scheme shows improved performance respect to other TDMA/CDMA-based schemes, while also not requiring additional  synchronization. 
%
%
\bibliographystyle{IEEEtran}
\bibliography{IEEEabrv,ICASSP21}

\end{document}

%% file: HetURA.tex
\begin{tikzpicture}[
  font=\small, >=stealth', line width=1.0pt,
  block/.style={rectangle, draw, minimum height=7mm, minimum width=10mm},
  block1/.style={rectangle, draw, minimum height=5mm, minimum width=7mm},
  mytrap/.style={
 	trapezium, trapezium angle=67.5, draw,inner xsep=0pt,outer sep=0pt,
 	minimum height=1.81mm, color=darkgray,text width=#1},
]
\def\d1{0.35}
\draw[dashed]  (0,2.5) -- (5.5,2.5) ;
\draw[dotted]  (0.5,-0.12) -- (5.5,-0.12) ;

\node (P1)  at (0,5) {};
\node (P2)  at (2,5) {};
\node (P3)  at (4,5) {};
\node (P4)  at (6,5) {};
\node (P5)  at (9,5){};
\node (P6)  at (12.5,5){};	
\node (P7)  at (15,5){};

\node (R1)  at (-1.5,5) {};
\node (R2)  at (-1.5,2.5) {};
\node (R3)  at (-1.5,-2.5) {};

\path[draw,decorate,decoration=brace,rotate =30] (R2) -- (R1)
node[midway,rotate=90,above,font=\small\sffamily] {Cluster 1};

\path[draw,decorate,decoration=brace] (R3) -- (R2)
node[midway,rotate=90,above,font=\small\sffamily] {Cluster 2};

\path[draw,decorate,decoration=brace] (P1) -- (P2)
node[midway,above,font=\small\sffamily]{fragmentation};

\node[below left=2.5*\d1 cm and 0*\d1 cm of P1, block] (wv1) {$\wv_{1i}$};
\node[below =8*\d1 cm of wv1, block] (wv2) {$\wv_{2i}$};


\path[draw,decorate,decoration=brace] (P2) -- (P3)
node[midway,above,font=\small\sffamily, text width=1.5cm]{tree encoding};

\node[below left=\d1 cm of P2, block1] (wv1i1) {\tiny $\wv_{1i1}$};
\node[below =0.1*\d1 cm of wv1i1] (wvi1dots) {\tiny $\vdots$};
\node[below =0.1*\d1 cm of wvi1dots, block1] (wv1iJ) {\tiny $\wv_{1iJ}$};

\node[below =1.6*\d1 cm of wv1iJ, block1] (wvo2i1) {\tiny $\wov_{2i1}$};
\node[below =0.1*\d1 cm of wvo2i1] (wvio2dots) {\tiny $\vdots$};
\node[below =0.1*\d1 cm of wvio2dots, block1] (wvo2iJ) {\tiny $\wov_{2iJ}$};

\node[below =1.6*\d1 cm of wvo2iJ, block1] (wvoo2i1) {\tiny $\woov_{2i1}$};
\node[below =0.1*\d1 cm of wvoo2i1] (wvioo2dots) {\tiny $\vdots$};
\node[below =0.1*\d1 cm of wvioo2dots, block1] (wvoo2iJ) {\tiny $\woov_{2iJ}$};


\draw [->,rounded corners,line width=0.01cm] (wv1.east) |- (wv1i1);
\draw [->,rounded corners,line width=0.01cm] (wv1.east) |- (wvi1dots);
\draw [->,rounded corners,line width=0.01cm] (wv1.east) |- (wv1iJ);

\draw [->,rounded corners,line width=0.01cm] (wv2.east) |- (wvo2i1) ;
\draw [->,rounded corners,line width=0.01cm] (wv2.east) |-  node[above, xshift=-8mm,font=\small\sffamily,text width=1cm] {bottom bits}  (wvio2dots);
\draw [->,rounded corners,line width=0.01cm] (wv2.east) |- (wvo2iJ);

\draw [->,rounded corners,line width=0.01cm] (wv2.east) |- (wvoo2i1);
\draw [->,rounded corners,line width=0.01cm] (wv2.east) |-  node[above, xshift=-8mm,font=\small\sffamily,text width=1cm] {top bits}   (wvioo2dots);
\draw [->,rounded corners,line width=0.01cm] (wv2.east) |- (wvoo2iJ);


\node[below left=\d1 cm of P3, block1] (t1) {\tiny $\Gv_{11}$};
\node[below =0.1*\d1 cm of t1] (t1dots) {\tiny $\vdots$};
\node[below =0.1*\d1 cm of t1dots, block1] (t1J) {\tiny $\Gv_{1J}$};

\node[below =1.6*\d1 cm of t1J, block1] (to1) {\tiny $\Gv_{11}$};
\node[below =0.1*\d1 cm of to1] (to1dots) {\tiny $\vdots$};
\node[below =0.1*\d1 cm of to1dots, block1] (toJ) {\tiny $\Gv_{1J}$};

\node[below =1.6*\d1 cm of toJ, block1] (t2) {\tiny $\Gv_{21}$};
\node[below =0.1*\d1 cm of t2] (t2dots) {\tiny $\vdots$};
\node[below =0.1*\d1 cm of t2dots, block1] (t2J) {\tiny $\Gv_{2J}$};


\draw [->,line width=0.01cm] (wv1i1.east) -- (t1);
\draw [->,line width=0.01cm] (wv1i1.east) -- (t1dots);
\draw [->,line width=0.01cm] (wv1i1.east) -- (t1J);

\draw [->,line width=0.01cm] (wv1iJ.east) -- (t1J);

\draw [->,line width=0.01cm] (wvo2i1.east) -- (to1);
\draw [->,line width=0.01cm] (wvo2i1.east) -- (to1dots);
\draw [->,line width=0.01cm] (wvo2i1.east) -- (toJ);

\draw [->,line width=0.01cm] (wvo2iJ.east) -- (toJ);

\draw [->,line width=0.01cm] (wvoo2i1.east) -- (t2);
\draw [->,line width=0.01cm] (wvoo2i1.east) -- (t2dots);
\draw [->,line width=0.01cm] (wvoo2i1.east) -- (t2J);

\draw [->,line width=0.01cm] (wvoo2iJ.east) -- (t2J);


\node[below left=\d1 cm of P4, block1] (cs1) {\tiny $\Av_{1}$};
\node[below =0.1*\d1 cm of cs1] (cs1dots) {\tiny $\vdots$};
\node[below =0.1*\d1 cm of cs1dots, block1] (csJ) {\tiny $\Av_{1}$};

\node[below =1.6*\d1 cm of csJ, block1] (cso1) {\tiny $\Av_1$};
\node[below =0.1*\d1 cm of cso1] (cso1dots) {\tiny $\vdots$};
\node[below =0.1*\d1 cm of cso1dots, block1] (csoJ) {\tiny $\Av_1$};

\node[below =1.6*\d1 cm of csoJ, block1] (cs2) {\tiny $\Av_2$};
\node[below =0.1*\d1 cm of cs2] (cs2dots) {\tiny $\vdots$};
\node[below =0.1*\d1 cm of cs2dots, block1] (cs2J) {\tiny $\Av_2$};


\draw [->,line width=0.01cm] (t1.east) -- node[above,font=\tiny\sffamily,mytrap=0.3cm] {be2i} (cs1);
\draw [->,line width=0.01cm] (t1J.east) -- node[above,font=\tiny\sffamily,mytrap=0.3cm] {be2i} (csJ);
\draw [->,line width=0.01cm] (to1.east) -- node[above,font=\tiny\sffamily,mytrap=0.3cm] {be2i} (cso1);
\draw [->,line width=0.01cm] (toJ.east) -- node[above,font=\tiny\sffamily,mytrap=0.3cm] {be2i} (csoJ);
\draw [->,line width=0.01cm] (t2.east) -- node[above,font=\tiny\sffamily,mytrap=0.3cm] {be2i} (cs2);
\draw [->,line width=0.01cm] (t2J.east) -- node[above,font=\tiny\sffamily,mytrap=0.3cm] {be2i} (cs2J);


\node[draw,rectangle, rounded corners, font=\small \sffamily] (MAC1) at (6.5,3) {concat};

\node[rectangle, rounded corners,draw, font=\small \sffamily] (MAC2) at (6.5,-0.1) {concat};

\node[draw,circle] (MAC) at (8,1.5) {$\sum$};



\draw [->,rounded corners,line width=0.01cm] (cs1.east) -- ++ (0.5,0cm)  --++  (MAC1);
\draw [->,rounded corners,line width=0.01cm] (csJ.east) -- ++ (MAC1);

\draw [->,rounded corners,line width=0.01cm] (cso1.east) -- ++ (0.5,0cm)  --++ (MAC2);
\draw [->,rounded corners,line width=0.01cm] (csoJ.east) -- ++ (0.5,0cm)  --++ (MAC2);

\draw [->,rounded corners,line width=0.01cm] (cs2.east) -- ++ (0.5,0cm)  --++ (MAC2);
\draw [->,rounded corners,line width=0.01cm] (cs2J.east) -- ++ (0.5,0cm)  --++ (MAC2);


\draw [->,rounded corners,line width=0.01cm] (MAC2.east) -|  node[below,xshift=10mm,font=\small\sffamily,text width=3cm] { $\xv_{2i}$}   (MAC.south);
\draw [->,rounded corners,line width=0.01cm] (MAC1.east) -|  node[above,xshift=10mm,font=\small\sffamily,text width=3cm] { $\xv_{1i}$}   (MAC.north);


\node[draw,circle,  minimum size=2mm,] (PLUS) at (11.1,3) {$+$};

\node[block,below  left = 9*\d1 cm and  8*\d1 cm of P6, rotate=90,minimum width=25mm] (csDecoder2) {CCS decoding -- $\Av_2$};
\node[block,below  left = 1.2*\d1 cm and  1.5*\d1 cm of P6, rotate=90,minimum width=25mm] (csDecoder1) {CCS decoding --  $\Av_1$};

edge[<-] (MAC);
\node[block,rotate=90,below  left = 10*\d1 cm and  1.5*\d1 cm of P7, minimum width=25mm] (treeDecoder2)  {Tree Decoder -- $\Gv_2$};
\node[block,rotate=90,below  left = 1*\d1 cm and  1.5*\d1 cm of P7, minimum width=25mm] (treeDecoder1) {Tree Decoder -- $\Gv_1$};

\path[draw,decorate,decoration=brace] (P3) -- (P4)
node[midway,above,font=\small\sffamily,text width=2cm]{Superposition +CS encode};

\path[draw,decorate,decoration=brace] (P4) -- (P5)
node[midway,above,font=\small\sffamily]{MAC channel};

\path[draw,decorate,decoration=brace] (P5) -- (P6)
node[midway,above,font=\small\sffamily,text width=2cm]{2-phase decoder};

\path[draw,decorate,decoration=brace] (P6) -- (P7)
node[midway,above,font=\small\sffamily,,text width=2cm]{tree decoding};
%

%

\draw [->,rounded corners,line width=0.01cm] (MAC1.east) -|  node[above,xshift=10mm,font=\small\sffamily,text width=3cm] { $\xv_{1i}$}   (MAC.north);

\draw [->,rounded corners,line width=0.01cm] ([yshift=-5.9pt]csDecoder2.south east) -|  node[above,xshift=3mm,yshift=5mm,font=\small \sffamily,text width=3cm] { $-\Phim_2 \mv_2$} (PLUS.south);

\draw [->,rounded corners,line width=0.01cm] (MAC.east) --  node[above,xshift=12mm,font=\small\sffamily,text width=3cm] { $\yv$}   ([yshift=-5.9pt]csDecoder2.north east);

\draw [->,rounded corners,line width=0.01cm] (MAC.east) --  node[above,xshift=12mm,font=\small\sffamily,text width=3cm] (yv) { $\yv$}   ([yshift=-5.9pt]csDecoder2.north east);

\draw [->,rounded corners,line width=0.01cm] (9.25,1.5) |-  (PLUS.west);

\draw [->,rounded corners,line width=0.01cm] (PLUS.east) -- ([yshift=1.25pt]csDecoder1.north);

%

\draw [->,rounded corners,line width=0.01cm] ([yshift=35+4pt]csDecoder1.south) -- node[above, xshift=2mm,font=\tiny \sffamily,text width=1cm] { $\wv_{1i1}$}   ([yshift=35+0pt]treeDecoder1.north);

\draw [->,rounded corners,line width=0.01cm] ([yshift=35-12-4 pt]csDecoder1.south) -- node[above, xshift=2mm,font=\tiny \sffamily,text width=1cm] { $\vdots$} node[below, xshift=2mm,font=\tiny \sffamily,text width=1cm] { $\wv_{1iJ}$}   ([yshift=35-16-4pt]treeDecoder1.north);


\draw [->,rounded corners,line width=0.01cm] ([yshift=-10+4pt]csDecoder1.south) -- node[above, xshift=2mm,font=\tiny \sffamily,text width=1cm] { $\wov_{2i1}$}   ([yshift=-10pt]treeDecoder1.north);

\draw [->,rounded corners,line width=0.01cm] ([yshift=-10-12-4 pt]csDecoder1.south) -- node[above, xshift=2mm,font=\tiny \sffamily,text width=1cm] { $\vdots$} node[below, xshift=2mm,font=\tiny \sffamily,text width=1cm] { $\wov_{2iJ}$}   ([yshift=-10-16-4pt]treeDecoder1.north);

\draw [->,rounded corners,line width=0.01cm] ([yshift=-8pt]csDecoder2.south) -- node[above, xshift=2mm,font=\tiny \sffamily,text width=1cm] { $\woov_{2i1}$}   ([yshift=0pt]treeDecoder2.north);


\draw [->,rounded corners,line width=0.01cm] ([yshift=-20-8 pt]csDecoder2.south) -- node[above, xshift=2mm,font=\tiny \sffamily,text width=1cm] { $\vdots$} node[below, xshift=2mm,font=\tiny \sffamily,text width=1cm] { $\woov_{2iJ}$}   ([yshift=-20-0pt]treeDecoder2.north);

\end{tikzpicture}

%% file: rate_saturation.tex
\begin{tikzpicture}
\definecolor{mycolor1}{rgb}{0.63529,0.07843,0.18431}%
\definecolor{mycolor2}{rgb}{0.00000,0.44706,0.74118}%
\definecolor{mycolor3}{rgb}{0.00000,0.49804,0.00000}%
\definecolor{mycolor4}{rgb}{0.87059,0.49020,0.00000}%
\definecolor{mycolor5}{rgb}{0.00000,0.44700,0.74100}%
\definecolor{mycolor6}{rgb}{0.74902,0.00000,0.74902}%

\begin{semilogyaxis}[%
font=\small,
width=7.5cm,
height=4cm,
scale only axis,
xmin=13,
xmax=23,
xlabel style={font=\color{white!15!black}},
xlabel={$P_1$ [dB]},
ymin=0,
ymax=1,
ylabel style={font=\color{white!15!black}},
ylabel={ $P_{\rm UE}$},
axis background/.style={fill=white},
xmajorgrids,
ymajorgrids,
legend style={legend cell align=left, align=left, draw=white!15!black, nodes={scale=0.75, transform shape}, at={(0.0,0.25)}, anchor=west, fill opacity=0.8}
]

\addplot [color=mycolor1, line width=1.5pt, mark=asterisk, mark options={solid, mycolor1}]
  table[row sep=crcr]{%
13 0.09333333333333338 \\
14 0.04488888888888887 \\
15 0.015555555555555545 \\
16 0.004444444444444473 \\
17 0.001777777777777767 \\
18 0.0004444444444444695 \\
19 0.0 \\
20 0.0 \\
21 0.0 \\
22 0.0 \\
23 0.0 \\
};
\addlegendentry{$R_1=0.03$}

\addplot [color=mycolor2, dashed, line width=1.5pt, mark=o, mark options={solid, mycolor2}]
  table[row sep=crcr]{%
13 0.7044444444444444 \\
14 0.48844444444444446 \\
15 0.2844444444444445 \\
16 0.0795555555555556 \\
17 0.01244444444444448 \\
18 0.0013333333333332975 \\
19 0.0 \\
20 0.0 \\
21 0.0 \\
22 0.0 \\
23 0.0 \\
};
\addlegendentry{$R_1=0.04$}

\addplot [color=mycolor3, dashdotted, line width=1.5pt, mark=diamond, mark options={solid, mycolor3}]
  table[row sep=crcr]{%
13 0.7044444444444444 \\
14 0.48844444444444446 \\
15 0.2844444444444445 \\
16 0.0795555555555556 \\
17 0.01244444444444448 \\
18 0.0013333333333332975 \\
19 0.0 \\
20 0.0 \\
21 0.0 \\
22 0.0 \\
23 0.0 \\
};
\addlegendentry{$R_1=0.05$}

\addplot [color=mycolor4, dashdotted, line width=1.5pt, mark=diamond, mark options={solid, mycolor4}]
  table[row sep=crcr]{%
13 0.904 \\
14 0.8528888888888889 \\
15 0.723111111111111 \\
16 0.4311111111111111 \\
17 0.1982222222222222 \\
18 0.04800000000000004 \\
19 0.007555555555555538 \\
20 0.0 \\
21 0.0 \\
22 0.0 \\
23 0.0 \\
};
\addlegendentry{$R_1=0.06$}

\addplot [color=mycolor5, dashdotted, line width=1.5pt, mark=diamond, mark options={solid, mycolor5}]
  table[row sep=crcr]{%
13 0.988 \\
14 0.964 \\
15 0.9342222222222222 \\
16 0.8502222222222222 \\
17 0.7306666666666667 \\
18 0.5262222222222221 \\
19 0.2466666666666667 \\
20 0.09199999999999997 \\
21 0.013777777777777778 \\
22 0.005333333333333301 \\
23 0.0 \\
};
\addlegendentry{$R_1=0.07$}

\addplot [color=mycolor6, dashdotted, line width=1.5pt, mark=diamond, mark options={solid, mycolor6}]
  table[row sep=crcr]{%
13 0.9933333333333333 \\
14 0.9862222222222222 \\
15 0.96 \\
16 0.9311111111111111 \\
17 0.8755555555555555 \\
18 0.6866666666666666 \\
19 0.4635555555555556 \\
20 0.28 \\
21 0.14800000000000002 \\
22 0.058666666666666645 \\
23 0.04533333333333334 \\
};
\addlegendentry{$R_1=0.075$}

\end{semilogyaxis}
\end{tikzpicture}%

%% file: Sim_one_layer_powerformance.tex

\begin{tikzpicture}
\definecolor{mycolor1}{rgb}{0.63529,0.07843,0.18431}%
\definecolor{mycolor2}{rgb}{0.00000,0.44706,0.74118}%
\definecolor{mycolor3}{rgb}{0.00000,0.49804,0.00000}%
\definecolor{mycolor4}{rgb}{0.87059,0.49020,0.00000}%
\definecolor{mycolor5}{rgb}{0.00000,0.44700,0.74100}%
\definecolor{mycolor6}{rgb}{0.74902,0.00000,0.74902}%

\begin{axis}[%
font=\small,
width=7cm,
height=5.5cm,
scale only axis,
every outer x axis line/.append style={white!15!black},
every x tick label/.append style={font=\color{white!15!black}},
xmin=17,
xmax=25,
xtick = {17,18,...,25},
xlabel={$P_1$ [dB]},
xmajorgrids,
every outer y axis line/.append style={white!15!black},
every y tick label/.append style={font=\color{white!15!black}},
ymin=0.06,
ymax=0.13,
ytick = {0.06,0.07,...,0.14},
ylabel={$R$},
ymajorgrids,
legend style={at={(0,1)},anchor=north west, draw=black,fill=white,legend cell align=left}
]

\addplot [color=mycolor2,densely dotted,line width=2.0pt,mark size=1.4pt,mark=o, mark options={solid}]
  table[row sep=crcr]{
17.0 0.07 \\
18.0 0.08 \\
19.0 0.09 \\
20.0 0.09 \\
21.0 0.09999999999999999 \\
22.0 0.09999999999999999 \\
23.0 0.09999999999999999 \\
23.020599913279625 0.09999999999999999 \\
24.0 0.09999999999999999 \\
24.020599913279625 0.10999999999999999 \\
25.0 0.10999999999999999 \\
25.020599913279625 0.10999999999999999 \\
26.020599913279625 0.10999999999999999 \\
27.020599913279625 0.10999999999999999 \\
28.020599913279625 0.12 \\
29.020599913279625 0.12 \\
30.020599913279625 0.12 \\
};
\addlegendentry{$R_1$};


\addplot [color=mycolor3, dashdotted, line width=1.5pt, mark=diamond, mark options={solid, mycolor3}]
  table[row sep=crcr]{%
17.0 0.09 \\
18.0 0.1 \\
19.0 0.11 \\
20.0 0.11 \\
21.0 0.11 \\
22.0 0.12 \\
23.0 0.12 \\
24.0 0.12 \\
};
\addlegendentry{$R_2$: our scheme}

\addplot [color=mycolor5, dashdotted, line width=1.5pt, mark=diamond, mark options={solid, mycolor5}]
  table[row sep=crcr]{%
17 0.065 \\
18 0.07400000000000001 \\
19 0.083 \\
20 0.092 \\
21 0.092 \\
22 0.101 \\
23 0.101 \\
24 0.101 \\
};
\addlegendentry{$R_2$: TDMA-scheme}
\end{axis}

\end{tikzpicture}%


%% file: SIC_CDMA.tex
\begin{tikzpicture}
\definecolor{mycolor1}{rgb}{0.63529,0.07843,0.18431}%
\definecolor{mycolor2}{rgb}{0.00000,0.44706,0.74118}%
\definecolor{mycolor3}{rgb}{0.00000,0.49804,0.00000}%
\definecolor{mycolor4}{rgb}{0.87059,0.49020,0.00000}%
\definecolor{mycolor5}{rgb}{0.00000,0.44700,0.74100}%
\definecolor{mycolor6}{rgb}{0.74902,0.00000,0.74902}%

\begin{semilogyaxis}[%
font=\small,
width=7.5cm,
height=4cm,
scale only axis,
xmin=18,
xmax=35,
xlabel style={font=\color{white!15!black}},
xlabel={$P_1$ [dB]},
ymin=0,
ymax=1/8,
ylabel style={font=\color{white!15!black}},
ylabel={ $R$},
axis background/.style={fill=white},
xmajorgrids,
ymajorgrids,
legend style={legend cell align=left, align=left, draw=white!15!black, nodes={scale=0.75, transform shape}, at={(0.21,0.25)}, anchor=west, fill opacity=0.8}
]

\addplot [color=mycolor1, line width=1.5pt, mark=asterisk, mark options={solid, mycolor1}]
  table[row sep=crcr]{%
18 0.0 \\
19 0.0 \\
20 0.0 \\
21 0.060000000000000005 \\
22 0.08000000000000002 \\
23 0.1 \\
24 0.11000000000000001 \\
25 0.11000000000000001 \\
26 0.11000000000000001 \\
27 0.11000000000000001 \\
28 0.11000000000000001 \\
};
\addlegendentry{$R_2-R_1$ for $M_1=15$}

\addplot [color=mycolor2, dashed, line width=1.5pt, mark=o, mark options={solid, mycolor2}]
  table[row sep=crcr]{%
18 0.0 \\
19 0.05 \\
20 0.060000000000000005 \\
21 0.08000000000000002 \\
22 0.1 \\
23 0.1 \\
24 0.11000000000000001 \\
25 0.11000000000000001 \\
26 0.11000000000000001 \\
27 0.11000000000000001 \\
28 0.11000000000000001 \\
};
\addlegendentry{$R_2-R_1$ for $M_1=10$}

\addplot [color=mycolor3, dashdotted, line width=1.5pt, mark=diamond, mark options={solid, mycolor3}]
  table[row sep=crcr]{%
28 0.0 \\
29 0.04 \\
30 0.05 \\
31 0.07 \\
32 0.08 \\
33 0.1 \\
34 0.11 \\
35 0.11 \\
36 0.11 \\
37 0.11 \\
38 0.11 \\
};
\addlegendentry{IAN prediction for $R_2-R_1$}

\end{semilogyaxis}
\end{tikzpicture}%